\documentclass[]{relaxed_system_lab}

\usepackage[utf8]{inputenc}
\usepackage[T1]{fontenc}
\usepackage{geometry}
\usepackage{amsmath,amssymb}
\usepackage{charter}
\usepackage{varwidth}
\usepackage{caption}
\usepackage{hyperref}
\usepackage{fontawesome5}

\usepackage[toc,page,header]{appendix}

\usepackage{minitoc}
\usepackage{cleveref}
\usepackage{subcaption}
\usepackage{booktabs}
\usepackage{graphicx}
\usepackage{pgfplots}
\usepackage{pgfplotstable}
\usepackage{xcolor}
\usepackage{CJKutf8}
\usetikzlibrary{patterns}
\usepackage{float}
\usepackage{bm}

\usepackage{soul}
\usepackage{algorithm}
\usepackage{algpseudocode}
\usepackage{parskip}

\usepackage{multirow}
\usepackage{url}
\usepackage{xspace}
\usepackage{tcolorbox}

\definecolor{lightgreen}{RGB}{144,238,144}
\definecolor{darkcyan}{HTML}{008B8B}

\definecolor{cvprblue}{rgb}{0.21,0.49,0.74}
\definecolor{fallbackgreen}{rgb}{130, 180, 102}
\definecolor{stopred}{rgb}{251, 225, 224}

\ifdefined\final
\usepackage[disable]{todonotes}
\else
\usepackage[textsize=tiny]{todonotes}
\fi
\newcommand{\name}{${{\mathcal{D}}}^2$ \textbf{SD}\xspace}

\newtcolorbox{promptbox}[1][]{
  enhanced,
  breakable,
  colback=promptboxlightgray,
  colframe=promptboxblue!30,
  arc=8pt,
  boxrule=0.5pt,
  left=12pt,
  right=12pt,
  top=8pt,
  bottom=8pt,
  fonttitle=\bfseries,
  fontupper=\linespread{1.2}\selectfont,
  title=#1
}

\title{\name: Accelerating Speculative Decoding with Dual Diffusion Draft Models}

\author{Liyuan Zhang$^1$, Jiarui Zhang$^{2,3,5}$, Jinwei Yao$^4$, Ran Yan$^{3,5}$, Yuchen Yang$^{2}$, Jiahao Zhang$^1$, Tongkai Yang$^5$, Yi Wu$^2$, Binhang Yuan$^3$}

\affiliation{$^1$Peking University, $^2$Tsinghua University, $^3$HKUST, $^4$UIUC, $^5$Ant Group}

\abstract{
Speculative decoding accelerates autoregressive large language model inference by drafting multiple tokens and verifying them in a single target-model forward pass. Recent diffusion-based drafters generate an entire block of tokens in parallel but usually commit to a single draft sequence per verification: once the first mismatch occurs, all subsequent draft tokens are discarded, resulting in a limited acceptance rate. Naively batching more draft candidate sequences only introduces a marginal improvement, as redundant or poorly placed branches increase the cost of drafting and verification without proportionally increasing the number of accepted tokens.
We propose \name, a \underline{\textbf{d}}ual \underline{\textbf{d}}iffusion draft \underline{\textbf{s}}peculative \underline{\textbf{d}}ecoding framework that organizes candidates into a confidence-guided prefix tree, where the first diffusion drafter generates a block along with per-position confidence scores that are used to identify the most likely rejection boundary and select the top-$K$ prefix ranges for recovery; the second variable-prefix diffusion drafter re-anchors at each selected prefix and proposes alternative continuations in one batched pass; the resulting shared-prefix candidates are jointly verified via cascade attention. Empirically, \name shows clear improvements over both the underlying diffusion approach and strong autoregressive speculative decoding baselines.
}

\begin{document}

\maketitle

\begin{center}
\vspace{-0.5em}
\href{https://github.com/catnanami/D2-SD}{%
  \faGithub\,\texttt{Code: https://github.com/catnanami/D2-SD}}
\vspace{0.5em}
\end{center}

\section{Introduction}
\label{sec:intro}

Autoregressive large language models (LLMs) achieve strong performance~\citep{liu2024deepseek,yang2025qwen3,zeng2026glm}, but the decoding computation remains inherently serial and memory-bandwidth bounded: each new token requires one forward pass through the original model. Speculative decoding alleviates this bottleneck by exploiting idle GPU compute to parallelize generation with a cheaper drafter, to propose multiple tokens in a single verification step over the original model (a.k.a. target model)~\citep{leviathan2023fast, chen2023accelerating}.

One fundamental challenge of speculative decoding lies in the sequential nature of its verification protocol: due to the \textit{longest-correct-prefix rule} (i.e., the speedup is determined by the longest prefix of the draft token sequence(s) accepted via verification), a single mismatch at an early position $i$ immediately invalidates all subsequent draft tokens in the sequence, regardless of their individual correctness. Note that this inherited issue creates a non-uniform importance across drafting positions---errors near the ``anchor'' (i.e., the last verified token) are catastrophic, while errors near the end of a block could be negligible. Consequently, simply increasing the drafting budget (i.e., generating more tokens) does not guarantee end-to-end speedup unless the additional compute is strategically allocated to minimize early rejections.

To address this, prefix tree-based speculative decoding (e.g., SpecInfer~\citep{miao2024specinfer}, Sequoia~\citep{chen2024sequoia}, and EAGLE-3~\citep{li2025eagle3}) organizes candidate sequences into branching structures to explore multiple continuations. While effective at increasing acceptance length, these methods are fundamentally limited by their autoregressive drafting phase. This sequential generation imposes a heavy ``drafting tax'' that scales poorly with the candidate budget: as the token budget increases from 4 to 16, the drafting overhead can balloon from 7ms to 24ms~\citep{chen2026dflash}. This serial bottleneck creates a diminishing performance boost where the time spent constructing a prefix tree with more candidate sequences often offsets the latency gains from its higher acceptance rate. Consequently, even with state-of-the-art verification kernels like tree-structured Flash Attention~\citep{pan2025fasttree,yaodeft}, the inherent latency of step-by-step drafting prevents these systems from effectively scaling to larger candidate batches in latency-sensitive applications.

Conversely, parallel drafting methods like DFlash~\citep{chen2026dflash} leverage a diffusion draft model to generate an entire block of tokens in a single forward pass, drastically reducing drafting latency. However, existing diffusion-style drafters typically apply a uniform computational budget across all positions within a block. This is inherently suboptimal: while early positions are often strongly constrained by the verified prefix, later positions face increasing predictive difficulty and require more diverse exploration. Our empirical study reveals that this computational mismatch leads to a ``scaling wall'' (Table~\ref{tab:gamma_saturation}): simply increasing the block size $\gamma$ yields diminishing or even negative returns once $\gamma \ge 24$, as an early error in a long, linear block renders the remaining extended budget useless. This bottleneck motivates our investigation into \textit{how to accelerate speculative decoding using diffusion-based draft model(s) with sophisticated construction and organization of draft sequences}.

\begin{table}[!ht]
\centering
%\small
\caption{DFlash acceptance length (\emph{tokens per forward pass}, TPF) on Qwen3-8B as the block size $\gamma$ grows; each row uses a decay-matched optimal training schedule for $\gamma$. TPF rises sharply from $\gamma{=}8$ to $\gamma{=}16$, saturates between $\gamma{=}16$ and $\gamma{=}24$, and \textit{decreases} at $\gamma{=}32$ ($\downarrow$). Simply enlarging the block thus exhibits diminishing and eventually negative returns---the drafted positions must instead be organized into a more effective structure.}
\label{tab:gamma_saturation}
\setlength{\tabcolsep}{8pt}
\begin{tabular}{c c c}
\toprule
Block size $\gamma$ & MATH-500 TPF & GSM8K TPF \\
\midrule
8  & 5.04 & 5.00 \\
16 & 6.05 & 5.95 \\
24 & 6.01 & 6.00 \\
32 & 5.85\,$\downarrow$ & 5.93\,$\downarrow$ \\
\bottomrule
\end{tabular}
\end{table}

Enabling parallel drafting with tree-structured budget allocation is non-trivial. The most intuitive baseline---generating $K$ independent candidates by sampling from the same diffusion drafter forward pass---fails to deliver meaningful gains. These candidates exhibit high error homogeneity: because they share the same underlying probability distribution, they tend to collapse onto the same high-confidence paths and collectively fail at the same difficult decision boundaries. As shown in our ablation (Table~\ref{tab:naive_branch}, full results in Section~\ref{sec:exp:second:resample}), adding $K{=}4$ naive branches to DFlash on GSM8K improves the acceptance length by only $0.37$ tokens (TPF $6.96 {\to} 7.33$), a marginal gain that is easily offset by the increased verification overhead, whereas \name lifts it by $2.25$ tokens (to $9.21$) under the same branching budget. This underscores that effective branching requires more than just increasing the sample count; it requires statistically informed recovery at likely rejection points using a model with a distinct inductive bias.

\begin{table}[!ht]
\centering
%\small
\caption{TPF on GSM8K (Qwen3-8B, $\gamma{=}16$, $K{=}4$): adding $K$ naive resampled branches to DFlash lifts TPF by only $0.37$, while \name raises it by $2.25$ under the same branching budget; full multi-dataset ablation in Section~\ref{sec:exp:second:resample}.}
\label{tab:naive_branch}
\setlength{\tabcolsep}{8pt}
\begin{tabular}{l c}
\toprule
Method & GSM8K TPF \\
\midrule
DFlash                  & 6.96 \\
\,+ $K$ naive samples   & 7.33 \\
\name                   & \textbf{9.21} \\
\bottomrule
\end{tabular}
\end{table}

Towards this end, we propose \name, a \underline{\textbf{d}}ual \underline{\textbf{d}}iffusion drafting paradigm to accelerate \underline{\textbf{s}}peculative \underline{\textbf{d}}ecoding, where our method differs by explicitly constructing a prefix-tree-like candidate set from the first diffusion drafter's confidence scores, so that re-drafting by the second specially trained drafter is performed only at likely rejection boundaries rather than through uniform, heuristic, or computationally expensive branching. Concretely, we make the following contributions:

\underline{\textbf{Contribution 1}}. We design a dual-draft paradigm: \underline{First}, a DFlash-style diffusion drafter generates an entire block in one parallel pass and produces per-position confidence scores. These confidence scores are then used to estimate the most likely rejection locations and to select the top-$K$ prefix ranges with the most promising recovery. \underline{Second}, we introduce a variable-prefix diffusion drafter that re-anchors at those selected prefix positions and proposes alternative suffixes in parallel, thereby converting a single diffusion draft into a structured set of shared-prefix candidates.

\underline{\textbf{Contribution 2}}. We implement this dual-draft paradigm with special training receipts and optimizations: To support this second stage, we design a dedicated training recipe that exposes the model to variable-length visible prefixes and applies an exponentially decaying training loss that emphasizes errors near the anchor. To enable efficient speculative decoding, we verify the resulting branches using cascade attention, allowing the method to reuse shared-prefix computation rather than replicating long KV prefixes across candidates.

\underline{\textbf{Contribution 3}}. We conduct a concrete empirical study to demonstrate clear improvements over strong speculative decoding baselines. For example, on Qwen3-8B model with a default configuration of ($\gamma{=}16$) and ($K{=}4$), \name increases the average greedy-decoding speedup from ($4.16\times$) to ($4.98\times$) relative to standard autoregressive decoding, while raising the average acceptance length from $5.31$ to $7.05$ when compared with DFlash.

\section{Preliminaries}
\label{sec:preliminaries}

\textbf{Speculative decoding speedup.}
Speculative decoding accelerates autoregressive inference of a target model $\mathcal{M}_t$ by introducing a lightweight draft model $\mathcal{M}_d$. In each decoding cycle, $\mathcal{M}_d$ proposes $\gamma$ candidate tokens, which $\mathcal{M}_t$ verifies in a single parallel forward pass and accepts up to the longest correct prefix.
Following~\cite{sadhukhan2024magicdec}, the average per-token latency of speculative decoding can be formulated as:
\begin{equation}
L \;=\; \frac{T_{\text{draft}} + T_{\text{verify}}}{\alpha},
\label{eq:spec_latency}
\end{equation}
where $T_{\text{draft}}$ is the time spent producing the draft block, $T_{\text{verify}}$ is the cost of one verification pass, and $\alpha \in [1, \gamma {+} 1]$ is the expected number of tokens committed per cycle (the mean acceptance length, including the bonus token returned by $\mathcal{M}_t$ at the rejection boundary). Letting $L_{\text{target}}$ denote the per-token latency of standard autoregressive decoding with $\mathcal{M}_t$, the resulting wall-clock speedup is
\begin{equation}
\eta \;=\; \frac{L_{\text{target}}}{L} \;=\; \frac{\alpha \cdot L_{\text{target}}}{T_{\text{draft}} + T_{\text{verify}}}.
\label{eq:spec_speedup}
\end{equation}
This expression makes the trade-off explicit: speedup grows either by raising the acceptance length $\alpha$ or by reducing the per-cycle drafting/verification overhead $T_{\text{draft}} {+}T_{\text{verify}}$. Block-diffusion drafters such as DFlash~\citep{chen2026dflash} drive $T_{\text{draft}}$ down by emitting a $\gamma$-token block in a single forward pass, but their reliance on a single draft sequence caps $\alpha$ well below $\gamma{+}1$ once an early mismatch occurs. \name targets exactly this gap: it raises $\alpha$ by re-drafting at the predicted rejection boundary while keeping $T_{\text{draft}}{+}T_{\text{verify}}$ nearly unchanged via a single batched second-draft pass and a shared-prefix joint verification.

\begin{figure}[t]
\centering
\includegraphics[width=\linewidth]{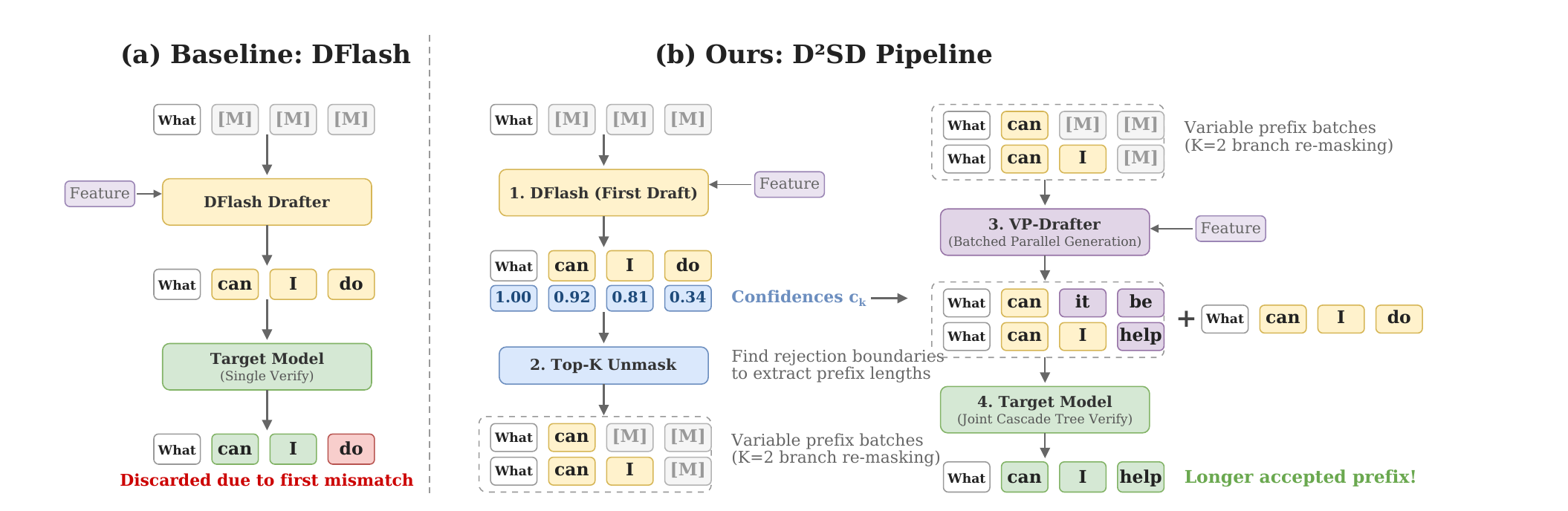}
\caption{\small \textbf{(a) Baseline DFlash} on a $\gamma = 4$ block. The DFlash drafter consumes the bonus-token anchor at position~$0$ and $\gamma{-}1$ mask positions, and predicts all mask positions in a single parallel pass conditioned on target hidden features via KV injection. The target model then verifies the drafted block by prefix matching, accepting tokens up to (but not including) the first mismatch and discarding the rest.
\textbf{(b) Our \name pipeline} on the same $\gamma{=}4$ block with $K {=} 2$ branches. \textbf{(1) DFlash (first draft)} runs as in (a), producing a first draft together with per-position confidences $c_k$. \textbf{(2) Top-$K$ Unmask} scores every prefix length $i$ with $r(i)$ (Eq.~\ref{eq:rate}) and selects the $K$ most-likely rejection boundaries, yielding $K$ variable-prefix batches that retain the chosen prefix and re-mask the remainder. \textbf{(3) VP-Drafter} fills in all $K$ batches in one batched forward pass, again conditioned on target hidden features. \textbf{(4) Target Model} jointly verifies the $K + 1$ candidates (the first-draft block plus $K$ second-draft branches) using shared-prefix cascade attention, and commits the longest accepted prefix across branches (see Section~\ref{sec:method:pipeline} for details).}
\label{fig:pipeline_contrast}
\end{figure}

\textbf{DFlash.} DFlash~\citep{chen2026dflash} is a block-diffusion speculative decoding framework that drafts $\gamma$ tokens in a single parallel forward pass; panel (a) of Figure~\ref{fig:pipeline_contrast} illustrates one decoding cycle as a drafter $\to$ target flow.
\textit{Drafting}: The drafter consumes a length-$\gamma$ input whose position $0$ holds the anchor---the bonus token returned by the previous verification step---and whose positions $1, \dots, \gamma {-} 1$ are all $[\text{MASK}]$. Conditioned on the target hidden features of the most recently accepted tokens---multi-layer target representations concatenated and projected by a learned $\mathrm{FC}$ layer, then injected into the Key and Value projections of every drafter layer---the drafter decodes all $\gamma {-} 1$ mask positions jointly in one parallel pass, with mask tokens attending bidirectionally to the target's representations. We highlight that the target model then verifies the drafted block by prefix matching, accepting tokens up to (but not including) the first mismatch and discarding the rest; the same forward pass also returns the next anchor (the target's prediction at the rejection boundary) and refreshed target hidden features that seed the next cycle.

\textbf{Limitation.} DFlash acceptance length is fundamentally bounded by the \textit{longest correct prefix}: once a mismatch occurs at some position, all subsequent draft tokens are discarded regardless of their quality. This observation motivates our dual-draft design, which re-drafts from the rejection boundary to recover these otherwise-lost positions.

\section{\name}
\label{sec:method}

We present \name (Dual-Draft Diffusion Speculative Decoding), designed to address the two failure modes identified in the introduction: the scaling wall of single-block diffusion drafting, where enlarging $\gamma$ yields diminishing or negative returns (Table~\ref{tab:gamma_saturation}); and the error homogeneity of resampling $K$ branches from the same DFlash forward pass. Our remedy has two components: rather than \emph{extend} the draft block, we re-anchor it at positions where the first draft is most likely to fail (Sections~\ref{sec:method:rejest}--\ref{sec:method:pipeline}); and rather than reuse DFlash for the second branch, we introduce a structurally distinct VP-Drafter (Variable-Prefix Drafter) trained to produce continuations from any prefix length within the block (Section~\ref{sec:method:training}), giving the cascade a different inductive bias from the first draft. Section~\ref{sec:method:rejest} explains how we use the first drafter's confidence to estimate where the rejection boundary is likely to occur; Section~\ref{sec:method:topk} turns this estimate into a top-$K$ set of shared-prefix branches; Section~\ref{sec:method:pipeline} describes the end-to-end inference pipeline; and Section~\ref{sec:method:training} presents the variable-length prefix training procedure.

\subsection{Estimating the Rejection Boundary from Drafter Confidence}
\label{sec:method:rejest}

The static, uniform allocation from DFlash does not match the verification rule: predictive difficulty is non-uniform across positions within a block. Enlarging $\gamma$ does not address this mismatch, since each position still receives the same per-position compute---the harder positions that bound acceptance are no better resourced than before, the rejection boundary does not move, and the added tokens are wasted (Table~\ref{tab:gamma_saturation}). \name replaces this static budget with a dynamic, difficulty-aware one: we re-anchor a second drafter at the predicted rejection boundary, directing an additional forward pass and a distinct inductive prior to the positions where the first draft was most likely to fail. Re-drafting too early wastes this extra capacity on positions that would have already been accepted, while re-drafting too late, on the other hand, leaves the actual rejection inside the kept prefix. Since the rejection boundary is unknown until verification, the second drafter must predict where the first draft will fail, and the remainder of this section shows that a lightweight drafter-internal signal can do so reliably.

We show that the first drafter's own per-position probability mass is sufficient. At each cycle DFlash produces a categorical $p_k(\cdot)$ at every block position $k = 1, \dots, \gamma{-}1$ and we greedy-sample $\hat{t}_k = \arg\max_v p_k(v)$; we define the \textit{confidence} of position $k$ as
\begin{equation}
c_k \;=\; p_k(\hat{t}_k) \;=\; \max_{v} p_k(v).
\label{eq:confidence}
\end{equation}
Figure~\ref{fig:confidence_accuracy} shows that $c_k$ is well-calibrated on GSM8K---the empirical accept rate tracks the diagonal across $[0, 1]$ and is slightly conservative in the high-confidence regime. Although this calibration property has been reported for autoregressive drafters~\citep{du2024glide,li2024eagle2}, to our knowledge it has not been verified for diffusion-based block drafters; our result shows that a target-conditioned block-diffusion drafter inherits the same property, letting us treat each $c_k$ as a reliable estimate of $\Pr(\hat{t}_k \text{ accepted} \mid \hat{t}_{<k} \text{ accepted})$. Assuming conditional independence given the prefix, the probability that exactly $i$ tokens are accepted ($L = i$) is
\begin{equation}
r(i) \;=\; \underbrace{\prod_{k=1}^{i} c_k}_{\text{first $i$ accepted}} \;\cdot\; \underbrace{(1 - c_{i+1})}_{\text{position $i{+}1$ rejected}}, \quad i = 0, 1, \dots, \gamma {-} 2.
\label{eq:rate}
\end{equation}
In other words, $r(i)$ turns the drafter's internal confidence into a posterior over rejection boundaries, telling us where the recovery budget should be spent.

\subsection{Top-$K$ Shared-Prefix Branch Selection}
\label{sec:method:topk}

Section~\ref{sec:method:rejest} produces a per-cycle posterior $r(i)$ over rejection boundaries; we now turn it into an actual branch set. As shown in the introduction (Table~\ref{tab:naive_branch}), drawing $K$ extra candidates from the same DFlash forward recovers only a small slice of the headroom, because all draws inherit the same per-position categorical and tend to collapse onto the same high-confidence paths. Effective branching, therefore, must answer two separate questions: where each branch should diverge from the first draft, and which model produces its continuation. We answer the first question here using $r(i)$; the second---a structurally distinct VP-Drafter that gives the cascade a different inductive bias---is deferred to Section~\ref{sec:method:training}.

\begin{figure}[!ht]
\centering
\begin{minipage}[c]{0.27\linewidth}%
  \centering
  \includegraphics[width=\linewidth]{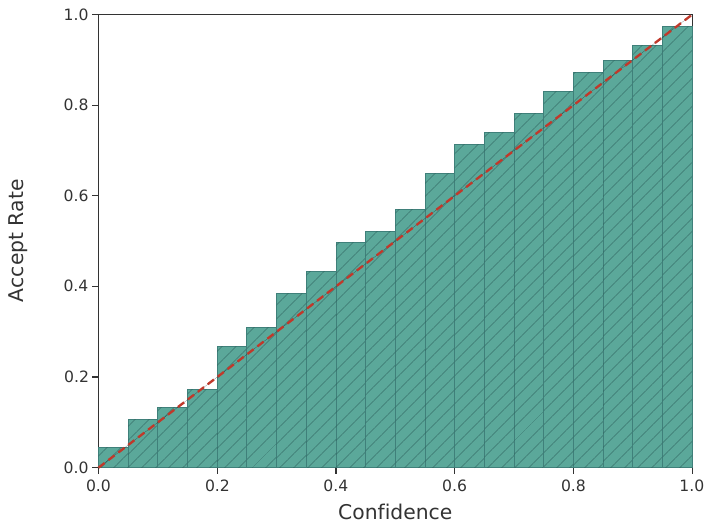}
  \subcaption{\small Confidence vs.\ accept rate.}
  \label{fig:confidence_accuracy}
\end{minipage}%
\hfill%
\begin{minipage}[c]{0.70\linewidth}%
  \centering
  \includegraphics[width=\linewidth]{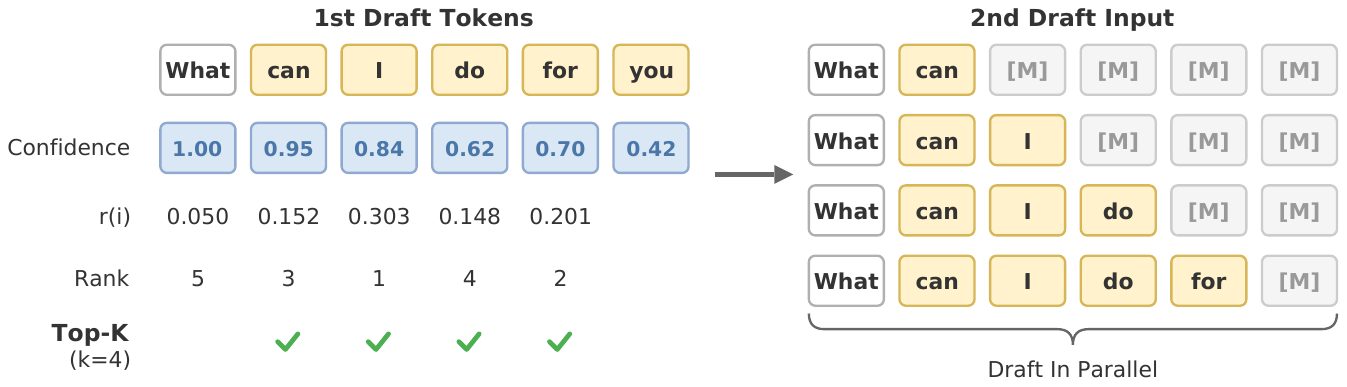}
  \subcaption{\small Top-$K$ unmask procedure.}
  \label{fig:topk_unmask}
\end{minipage}
\caption{\small \textbf{(a)} DFlash drafter confidence (Eq.~\ref{eq:confidence}) versus empirical accept rate. Each bar is the mean accept rate of draft tokens whose confidence falls into the corresponding bin; the red dashed line is $y = x$. The drafter is near-diagonally calibrated and slightly conservative in the high-confidence regime.
\textbf{(b)} Top-$K$ unmask with $K = 4$ on a $\gamma = 6$ block: we score every candidate prefix length with $r(i)$ in Eq.~(\ref{eq:rate}) and select the four highest (green checks). Each selected prefix yields one second-draft input that retains the prefix and re-masks the rest; all branches are batched into one VP-Drafter forward pass.}
\label{fig:unmask_combined}
\end{figure}

Committing to the single $\arg\max_i r(i)$ would forfeit most of the recovery opportunity, because $r(i)$ is typically diffuse---several prefix lengths receive comparable mass. We therefore retain the top-$K$ prefix lengths,
\begin{equation}
\mathcal{S} \;=\; \operatorname*{Top\text{-K}}_{i \in \{0, \dots, \gamma - 2\}}\; r(i),
\label{eq:topk}
\end{equation}
and launch one second-draft branch from each. Branch $i \in \mathcal{S}$ retains the anchor and the first $i$ DFlash tokens as its visible prefix and re-drafts the remaining $\gamma {-} 1 {-} i$ positions with the VP-Drafter (Figure~\ref{fig:topk_unmask}). By construction, every branch shares its first $i {+} 1$ tokens with the corresponding DFlash prefix, so the resulting candidate set has exactly the shared-prefix structure that cascade-style joint verification (Section~\ref{sec:method:pipeline}) can exploit---inheriting the verification-amortization benefit of tree-based speculative decoding~\citep{miao2024specinfer,chen2024sequoia,li2024eagle2,cai2024medusa} without paying their autoregressive drafting tax.

\FloatBarrier
\subsection{End-to-End Inference Pipeline}
\label{sec:method:pipeline}

A full \name decoding cycle combines the DFlash first draft, the VP-Drafter second draft, and joint verification. Each cycle consumes the verified KV caches of the target and the two drafters, along with the bonus token from the previous cycle, and emits a new batch of accepted tokens. Panel (b) of Figure~\ref{fig:pipeline_contrast} illustrates one such cycle on a $\gamma = 4$ block; the four stages are described below.

\textbf{(i) First draft (DFlash).} As in standard DFlash, the drafter builds a length-$\gamma$ masked block: position $0$ holds the bonus token (the anchor) and positions $1, \dots, \gamma {-} 1$ are $[\text{MASK}]$. Conditioned on the target hidden features from the previous cycle, DFlash predicts all $\gamma - 1$ mask positions in one pass, yielding tokens $\hat{t}_1, \dots, \hat{t}_{\gamma - 1}$ with confidences $c_1, \dots, c_{\gamma - 1}$.

\textbf{(ii) Top-$K$ unmask.} From the confidences we compute $r(i)$ (Eq.~\ref{eq:rate}) for each $i \in \{0, \dots, \gamma {-} 2\}$ and pick the top-$K$ prefix lengths $\mathcal{S}$ (Eq.~\ref{eq:topk}); each $i \in \mathcal{S}$ becomes a re-anchoring position.

\textbf{(iii) Second draft (VP-Drafter, batched).} For every $i \in \mathcal{S}$, the second-draft input retains the anchor and the first $i$ DFlash tokens as the visible prefix and re-masks positions $i {+} 1, \dots, \gamma {-} 1$. The $K$ branches are stacked into a single batch and processed by the VP-Drafter in one parallel pass, reusing the same target hidden features as DFlash. Each output branch agrees with DFlash on its first $i {+} 1$ tokens and proposes alternative continuations for the remainder.

\textbf{(iv) Joint verification and acceptance.} Together with the original DFlash branch, the target model jointly verifies the $K {+} 1$ candidates in a single forward pass. We then run a longest-accepted-prefix search across branches and commit the longest accepted prefix; the target's prediction at the rejection position becomes the next anchor (the bonus token).

\subsection{Training the VP-Drafter}
\label{sec:method:training}

The VP-Drafter shares its architecture with DFlash---a lightweight Qwen-based block-diffusion model with KV-injected target features---but uses a different masking schedule designed to support \emph{any} prefix length within the block.

 At inference, the unmask procedure of Section~\ref{sec:method:topk} draws $K$ different prefix lengths from $\{0, \dots, \gamma {-} 2\}$, and the second drafter must produce sensible continuations from each of them. DFlash, by contrast, is trained with a single fixed anchor at position $0$; reusing it for the second stage would force extrapolation to unseen prefix lengths, sharply degrading acceptance as the prefix grows.

 For each training example we sample a prefix length $l$ from a truncated geometric prior
\begin{equation}
\Pr(l = j) \;\propto\; \beta^{j}, \quad j = 0, 1, \dots, \gamma {-} 2,
\label{eq:vp_prior}
\end{equation}
with $\beta \in (0, 1)$ biasing toward shorter prefixes. The first $l {+} 1$ positions are filled with ground-truth tokens and positions $l {+} 1, \dots, \gamma {-} 1$ are masked. The drafter then predicts the masked positions in parallel, conditioned on target hidden features extracted from the same teacher pass used by DFlash. Because verification accepts tokens sequentially from the anchor outward, errors on positions closer to the anchor are strictly more costly than errors on later positions. We therefore train with an exponentially decayed cross-entropy that weights each masked position by its distance from the anchor,
\begin{equation}
\mathcal{L}_{\text{VP}} \;=\; -\,\frac{\sum_{k = l + 1}^{\gamma - 1} w_k \, \log p_k\!\left(t_k^{\star}\right)}{\sum_{k = l + 1}^{\gamma - 1} w_k}, \qquad w_k \;=\; \exp\!\left(-\frac{k - l - 1}{\tau}\right),
\label{eq:vp_loss}
\end{equation}
with decay rate $\tau$ and the expectation taken over data and the prior in Eq.~(\ref{eq:vp_prior}).

\section{Experiments}
\label{sec:experiments}

\subsection{Setup}
\label{sec:exp:setup}

\textbf{Models.}
We use Qwen3-8B and GPT-OSS-20B as the target model throughout. All experiments are conducted on NVIDIA H200 GPUs unless otherwise noted.

\textbf{Benchmarks.}
We evaluate on eight tasks spanning three categories: \textbf{Math}---GSM8K~\citep{cobbe2021gsm8k}, MATH~\citep{hendrycks2021math}, and AIME25~\citep{maa2025aime}; \textbf{Code}---HumanEval~\citep{chen2021humaneval}, MBPP~\citep{austin2021mbpp}, and LiveCodeBench~\citep{jain2024livecodebench}; \textbf{Chat}---MT-Bench~\citep{zheng2023mtbench} and Alpaca~\citep{taori2023alpaca}.

\textbf{Training data.}
All the EAGLE-3, DFlash and the VP-Drafter are trained on the full PerfectBlend\footnote{\url{https://huggingface.co/datasets/mlabonne/open-perfectblend}} dataset~\citep{xu2024perfectblendredefiningrlhf}, which aggregates diverse instruction--response pairs across mathematics, code, chat, and instruction following domains, unless otherwise noted. To ensure the draft models are aligned with the target distribution, all training responses are generated by the target model itself.

\textbf{Metrics.}
We measure the \textit{mean acceptance length} $\alpha$, defined as the average number of tokens accepted per decoding cycle, and the \textit{wall-clock speedup}, defined as the ratio of the autoregressive baseline's per-token decode latency to that of the speculative method.

\textbf{Implementation.}
All models are served in BF16 precision with FlashAttention-2~\citep{dao2024flashattention2}. Cascade attention is implemented with flashinfer~\citep{ye2025flashinfer} kernels and pre-warmed CUDA graphs for both the VP-Drafter and the target verification pass. The default configuration uses second-draft branch sizes $\gamma {=} 16$ and $K {=} 4$. We use HuggingFace Transformers as the model backbone and train the models on the SpecForge\footnote{\url{https://github.com/sgl-project/SpecForge}}.

\subsection{End-to-End Results}
\label{sec:exp:main}

Tables~\ref{tab:main} and~\ref{tab:gpt_oss_20b_dflash_ckpt} report wall-clock
speedup and mean acceptance length for DFlash, EAGLE-3~\citep{li2025eagle3},
and \name ($K{=}4$, $\gamma{=}16$) on Qwen3-8B and GPT-OSS-20B under greedy
($T{=}0$) and multinomial ($T{=}1$) decoding. Across both targets and both
decoding regimes, \name consistently improves over DFlash on both metrics, with
the largest gains on math and code where tighter rejection boundaries give the
second drafter more headroom to recover, and smaller gains on open-ended chat
where the boundary posterior $r(i)$ is more diffuse. Against the autoregressive
EAGLE-3, \name wins on wall-clock speedup on every task even when EAGLE-3
occasionally reaches a slightly longer $\alpha$, confirming that the latency
advantage of parallel block drafting outweighs EAGLE-3's marginal acceptance
edge. The advantage is amplified on the larger GPT-OSS-20B target, where
heavier per-step verification makes each token recovered by re-anchoring
translate into a larger wall-clock saving; under $T{=}1$ all methods degrade as
draft--target agreement weakens, but \name retains a consistent lead.

\FloatBarrier
\begin{table}[!ht]
\centering
\caption{Wall-clock speedup over autoregressive decoding and mean acceptance length ($\alpha$) on Qwen3-8B ($\gamma{=}16$, $K{=}4$) under greedy ($T{=}0$) and sampling ($T{=}1$) decoding. EAGLE-3~\citep{li2025eagle3}, DFlash, and \textsc{D$^2$SD} are measured on the same public Qwen3-8B checkpoint with matched hardware and decoding hyperparameters; we report end-to-end wall-clock speedup and mean acceptance length from our runs at both temperatures. Row-wise averages are arithmetic means over the eight datasets.}
\label{tab:main}
\small
\setlength{\tabcolsep}{4pt}
\resizebox{\textwidth}{!}{%
\begin{tabular}{l ccc ccc ccc ccc}
\toprule
& \multicolumn{6}{c}{\textbf{Greedy} ($T{=}0$)} & \multicolumn{6}{c}{\textbf{Sampling} ($T{=}1$)} \\
\cmidrule(lr){2-7} \cmidrule(lr){8-13}
& \multicolumn{3}{c}{Speedup} & \multicolumn{3}{c}{Acc.\ Len.} & \multicolumn{3}{c}{Speedup} & \multicolumn{3}{c}{Acc.\ Len.} \\
\cmidrule(lr){2-4} \cmidrule(lr){5-7} \cmidrule(lr){8-10} \cmidrule(lr){11-13}
\textbf{Dataset} & DFlash & EAGLE-3 & \textsc{D$^2$SD} & DFlash & EAGLE-3 & \textsc{D$^2$SD} & DFlash & EAGLE-3 & \textsc{D$^2$SD} & DFlash & EAGLE-3 & \textsc{D$^2$SD} \\
\midrule
GSM8K         & 5.51$\times$ & 4.14$\times$ & \textbf{6.54}$\times$ & 6.96 & 6.36 & \textbf{9.21} & 4.62$\times$ & 3.72$\times$ & \textbf{5.32}$\times$ & 5.84 & 5.93 & \textbf{7.38} \\
MATH          & 5.11$\times$ & 4.11$\times$ & \textbf{6.03}$\times$ & 6.55 & 6.44 & \textbf{8.56} & 4.25$\times$ & 3.81$\times$ & \textbf{4.90}$\times$ & 5.38 & 6.07 & \textbf{6.81} \\
AIME25        & 4.32$\times$ & 4.03$\times$ & \textbf{5.10}$\times$ & 5.49 & 6.33 & \textbf{7.26} & 3.55$\times$ & 3.57$\times$ & \textbf{4.11}$\times$ & 4.53 & 5.73 & \textbf{5.75} \\
\midrule
HumanEval     & 4.39$\times$ & 3.85$\times$ & \textbf{5.37}$\times$ & 5.50 & 5.92 & \textbf{7.47} & 3.49$\times$ & 3.47$\times$ & \textbf{4.05}$\times$ & 4.35 & \textbf{5.59} & 5.50 \\
MBPP          & 4.07$\times$ & 3.92$\times$ & \textbf{5.03}$\times$ & 5.09 & 6.03 & \textbf{6.96} & 3.25$\times$ & 3.50$\times$ & \textbf{4.13}$\times$ & 4.03 & \textbf{5.57} & 4.88 \\
LiveCodeBench & 3.88$\times$ & 3.70$\times$ & \textbf{4.62}$\times$ & 4.86 & 5.93 & \textbf{6.44} & 3.06$\times$ & 3.10$\times$ & \textbf{3.53}$\times$ & 3.85 & \textbf{5.01} & 4.90 \\
\midrule
MT-Bench      & 3.23$\times$ & 3.35$\times$ & \textbf{3.80}$\times$ & 4.41 & 5.22 & \textbf{5.75} & 2.69$\times$ & 2.79$\times$ & \textbf{3.09}$\times$ & 3.52 & \textbf{4.40} & 4.36 \\
Alpaca        & 2.79$\times$ & 3.30$\times$ & \textbf{3.38}$\times$ & 3.59 & \textbf{5.09} & 4.74 & 2.50$\times$ & 2.73$\times$ & \textbf{2.97}$\times$ & 3.17 & \textbf{4.33} & 4.02 \\
\midrule
\textbf{Average} & 4.16$\times$ & 3.80$\times$ & \textbf{4.98}$\times$ & 5.31 & 5.91 & \textbf{7.05} & 3.43$\times$ & 3.34$\times$ & \textbf{4.01}$\times$ & 4.33 & 5.33 & \textbf{5.45} \\
\bottomrule
\end{tabular}%
}
\end{table}

\FloatBarrier
\begin{table}[!ht]
\centering
\caption[GPT-OSS-20B with Hugging Face DFlash drafter checkpoint]{GPT-OSS-20B with the public DFlash drafter \texttt{z-lab/gpt-oss-20b-DFlash} as draft\,1 (\protect\url{https://huggingface.co/z-lab/gpt-oss-20b-DFlash}): wall-clock speedup over autoregressive decoding and mean acceptance length ($\alpha$) ($\gamma{=}16$, $K{=}4$) under greedy ($T{=}0$) and sampling ($T{=}1$). DFlash and \textsc{D$^2$SD} use reruns with this checkpoint (same hardware and decoding hyperparameters otherwise). Row-wise averages are arithmetic means over the eight datasets.}
\label{tab:gpt_oss_20b_dflash_ckpt}
\small
\setlength{\tabcolsep}{4pt}
\resizebox{\textwidth}{!}{%
\begin{tabular}{l ccc ccc ccc ccc}
\toprule
& \multicolumn{6}{c}{\textbf{Greedy} ($T{=}0$)} & \multicolumn{6}{c}{\textbf{Sampling} ($T{=}1$)} \\
\cmidrule(lr){2-7} \cmidrule(lr){8-13}
& \multicolumn{3}{c}{Speedup} & \multicolumn{3}{c}{Acc.\ Len.} & \multicolumn{3}{c}{Speedup} & \multicolumn{3}{c}{Acc.\ Len.} \\
\cmidrule(lr){2-4} \cmidrule(lr){5-7} \cmidrule(lr){8-10} \cmidrule(lr){11-13}
\textbf{Dataset} & DFlash & EAGLE-3 & \textsc{D$^2$SD} & DFlash & EAGLE-3 & \textsc{D$^2$SD} & DFlash & EAGLE-3 & \textsc{D$^2$SD} & DFlash & EAGLE-3 & \textsc{D$^2$SD} \\
\midrule
GSM8K         & 2.83$\times$ & 2.32$\times$ & \textbf{6.20}$\times$ & 3.58 & 2.81 & \textbf{8.02} & 1.65$\times$ & 1.17$\times$ & \textbf{1.94}$\times$ & 2.00 & 1.45 & \textbf{2.49} \\
MATH          & 2.48$\times$ & 2.33$\times$ & \textbf{5.43}$\times$ & 3.01 & 2.84 & \textbf{7.03} & 1.90$\times$ & 1.18$\times$ & \textbf{2.27}$\times$ & 2.29 & 1.46 & \textbf{2.91} \\
AIME25        & 1.83$\times$ & 2.26$\times$ & \textbf{5.18}$\times$ & 2.14 & 2.74 & \textbf{6.51} & 1.89$\times$ & 1.16$\times$ & \textbf{2.07}$\times$ & 2.21 & 1.44 & \textbf{2.54} \\
\midrule
HumanEval     & 3.50$\times$ & 2.44$\times$ & \textbf{6.33}$\times$ & 4.09 & 2.99 & \textbf{7.86} & 1.71$\times$ & 1.17$\times$ & \textbf{1.93}$\times$ & 1.97 & 1.44 & \textbf{2.37} \\
MBPP          & 5.33$\times$ & 2.44$\times$ & \textbf{8.87}$\times$ & 6.21 & 2.96 & \textbf{11.94} & 1.65$\times$ & 1.18$\times$ & \textbf{1.69}$\times$ & 1.93 & 1.45 & \textbf{2.23} \\
LiveCodeBench & 4.72$\times$ & 2.42$\times$ & \textbf{6.64}$\times$ & 5.54 & 2.93 & \textbf{9.02} & \textbf{1.68}$\times$ & 1.19$\times$ & 1.62$\times$ & 1.97 & 1.45 & \textbf{2.17} \\
\midrule
MT-Bench      & 2.65$\times$ & 2.23$\times$ & \textbf{2.67}$\times$ & 3.01 & 2.70 & \textbf{3.44} & \textbf{1.60}$\times$ & 1.15$\times$ & 1.59$\times$ & 1.80 & 1.44 & \textbf{2.04} \\
Alpaca        & 4.88$\times$ & 2.36$\times$ & \textbf{7.87}$\times$ & 5.49 & 2.87 & \textbf{10.37} & \textbf{1.58}$\times$ & 1.14$\times$ & \textbf{1.58}$\times$ & 1.78 & 1.41 & \textbf{2.05} \\
\midrule
\textbf{Average} & 3.53$\times$ & 2.35$\times$ & \textbf{6.15}$\times$ & 4.13 & 2.85 & \textbf{8.02} & 1.71$\times$ & 1.17$\times$ & \textbf{1.84}$\times$ & 1.99 & 1.44 & \textbf{2.35} \\
\bottomrule
\end{tabular}%
}
\end{table}

\subsection{Ablation Study of \name}

To justify the design of \name, we conduct the following ablation studies:

\underline{\textbf{Ablation 1.}} \textit{Why simply augmenting DFlash with $K$ resampled branches fails?}
\label{sec:exp:second:resample}
A natural baseline for the VP-Drafter is to skip the second model and let DFlash itself produce extra candidates: we keep the original argmax draft and append $K{=}4$ branches drawn as per-position $T{=}1$ multinomial samples from the same drafter forward, verifying all $K{+}1$ jointly via cascade attention. This matches \name's branching budget, so any difference reflects the source of branch diversity rather than verifier capacity.
The resampled branches recover only a small slice of the headroom---average $\alpha$ rises from $5.75$ to $6.08$ and speedup from $4.48\times$ to $5.06\times$, well short of the $\alpha = 7.62$ that \name attains. The reason is structural: all $K{+}1$ branches share one drafter forward and therefore the same per-position categorical $p_k$. At confident positions every sample collapses to the kept argmax, while at the diffuse positions that determine the rejection boundary the $K$ i.i.d.\ draws cannot reach probability mass outside $p_k$'s support. Resampling the same drafter thus cannot inject information it does not already carry; closing the remaining gap requires a model with a different inductive bias---the VP-Drafter, trained for variable-length prefixes (Section~\ref{sec:method:training}).

\begin{table}[!ht]
\centering
%\small
\caption{Augmenting argmax DFlash with $K{=}4$ resampled branches yields only modest gains over single-branch DFlash, while \name improves both metrics substantially under the same branching budget. Greedy target decoding ($T{=}0$); each cell is speedup $\times$ / mean acceptance length $\alpha$.}
\label{tab:dflash_resample}
\setlength{\tabcolsep}{6pt}
\begin{tabular}{l c c c}
\toprule
Dataset & DFlash & $+K$ samples & \name \\
\midrule
GSM8K    & 5.51$\times$/6.96 & 6.18$\times$/7.33 & \textbf{6.54$\times$/9.21} \\
MATH     & 5.11$\times$/6.55 & 5.67$\times$/6.82 & \textbf{6.03$\times$/8.56} \\
MBPP     & 4.07$\times$/5.09 & 4.66$\times$/5.47 & \textbf{5.03$\times$/6.96} \\
MT-Bench & 3.23$\times$/4.41 & 3.74$\times$/4.70 & \textbf{3.80$\times$/5.75} \\
\midrule
Avg.     & 4.48$\times$/5.75 & 5.06$\times$/6.08 & \textbf{5.35$\times$/7.62} \\
\bottomrule
\end{tabular}
\end{table}

\underline{\textbf{Ablation 2.}} \textit{Why not directly reusing DFlash as the second drafter?}
\label{sec:exp:second:reuse}
The complementary baseline keeps \name's full pipeline---top-$K$ confidence unmask, variable-prefix re-anchoring, and joint cascade verification---but reuses the original DFlash model as the second drafter, so the only change from \name is the inductive bias of the second draft. Because DFlash is trained with a fixed anchor at position $0$, applying it to the variable-length prefixes the cascade selects forces extrapolation to settings it has never seen.

The cascade structure clearly helps: average $\alpha$ rises to $6.53$ ($+14\%$ over single-branch DFlash, against only $+6\%$ for the naive resampling baseline), confirming that re-anchoring at the predicted rejection boundary places branches more effectively than uniform sampling. Wall-clock speedup, however, rises only modestly to $4.69\times$ ($+5\%$), since the second drafter is now a full forward pass rather than a logits view of the first. Full \name still leads by $1.09$ tokens of $\alpha$ ($+17\%$) and $0.66\times$ of speedup ($+14\%$)---precisely the gap the variable-prefix training of Eq.~(\ref{eq:vp_loss}) buys by calibrating the second drafter to exactly the longer prefixes the cascade most relies on.

\begin{table}[!ht]
\centering
\small
\caption{Reusing DFlash as the second drafter inside \name's cascade pipeline ($K{=}4$). Confidence-guided branch placement raises $\alpha$ over single-branch DFlash, but the lack of variable-prefix training keeps both metrics below full \name. Each cell is speedup $\times$ / mean acceptance length $\alpha$.}
\label{tab:dflash_reuse}
\setlength{\tabcolsep}{6pt}
\begin{tabular}{l c c c}
\toprule
Dataset & DFlash & DFlash$\to$DFlash & \name \\
\midrule
GSM8K    & 5.51$\times$/6.96 & 5.76$\times$/7.95 & \textbf{6.54$\times$/9.21} \\
MATH     & 5.11$\times$/6.55 & 5.28$\times$/7.35 & \textbf{6.03$\times$/8.56} \\
MBPP     & 4.07$\times$/5.09 & 4.36$\times$/5.90 & \textbf{5.03$\times$/6.96} \\
MT-Bench & 3.23$\times$/4.41 & 3.35$\times$/4.91 & \textbf{3.80$\times$/5.75} \\
\midrule
Avg.     & 4.48$\times$/5.75 & 4.69$\times$/6.53 & \textbf{5.35$\times$/7.62} \\
\bottomrule
\end{tabular}
\end{table}

\underline{\textbf{Ablation 3.}} \textit{Why not using more cascaded levels?}
A natural question is whether cascading the same idea once more buys further headroom. We test this by stacking a third VP-Drafter level on top of \name: for every $i \in \mathcal{S}$, we re-apply the rejection-boundary estimator (Eq.~\ref{eq:rate}) to that branch's second-draft tail, take the single top-$1$ location $j_i$, and dispatch a third VP-Drafter forward anchored there. This yields up to $2K {+} 1$ shared-prefix candidates that the target jointly verifies in a single pass; reusing the VP-Drafter is natural since its variable-prefix training (Section~\ref{sec:method:training}) already covers any prefix length in $\{1, \dots, \gamma {-} 2\}$.

The third level buys acceptance length but loses wall-clock speedup (Table~\ref{tab:third_draft}): averaged across the four datasets, $\alpha$ rises from $7.62$ to $8.24$ ($+8\%$) while speedup regresses from $5.35\times$ to $5.13\times$ ($-4\%$). The asymmetry is structural: the third pass operates on the residual probability mass the second did not already cover, so its marginal acceptance contribution ($0.62$ tokens) is only about a third of the second pass's gain over single-branch DFlash ($1.87$ tokens). Meanwhile both halves of the per-cycle cost grow---an extra VP-Drafter forward in $T_{\text{draft}}$ and roughly double the cascade-attention work in $T_{\text{verify}}$---inflating $T_{\text{draft}} {+} T_{\text{verify}}$ by $\sim$$13\%$ against only $\sim$$8\%$ acceptance gain. Stacking deeper would only sharpen this asymmetry, so \name commits to a single re-anchoring stage.

\begin{table}[!ht]
\centering
%\small
\caption{Stacking a third VP-Drafter level on top of \name ($K{=}4$). Every second-draft branch is re-anchored once at its own top-$1$ predicted rejection boundary, producing up to $2K {+} 1 = 9$ shared-prefix candidates per cycle. Mean acceptance length improves modestly, but the extra drafting and verification cost outweighs the marginal recovery, regressing wall-clock speedup. Each cell is speedup $\times$ / mean acceptance length $\alpha$.}
\label{tab:third_draft}
\setlength{\tabcolsep}{6pt}
\begin{tabular}{l c c}
\toprule
Dataset & \name & $+$ 3rd draft \\
\midrule
GSM8K    & \textbf{6.54}$\times$/9.21 & 6.25$\times$/\textbf{9.89} \\
MATH     & \textbf{6.03}$\times$/8.56 & 5.79$\times$/\textbf{9.27} \\
MBPP     & \textbf{5.03}$\times$/6.96 & 4.85$\times$/\textbf{7.58} \\
MT-Bench & \textbf{3.80}$\times$/5.75 & 3.63$\times$/\textbf{6.20} \\
\midrule
Avg.     & \textbf{5.35}$\times$/7.62 & 5.13$\times$/\textbf{8.24} \\
\bottomrule
\end{tabular}
\end{table}

\section{Related Work}
\label{sec:related}

\textbf{Speculative Decoding.} Speculative decoding~\citep{leviathan2023fast,chen2023accelerating} accelerates autoregressive LLM inference by employing a lightweight drafter to propose candidate tokens that are then verified once by the target model, yielding lossless acceleration through rejection sampling. The core question is how to obtain a drafter that is both cheap to run and well aligned with the target distribution. Medusa~\citep{cai2024medusa} removes the external drafter altogether by attaching multiple prediction heads to the target LLM. The EAGLE series~\citep{li2024eagle,li2024eagle2,li2025eagle3} instead reuses the target's frozen hidden states as drafter input and progressively refines draft quality, currently defining the autoregressive state of the art. Complementary directions include lookahead decoding via Jacobi trajectories~\citep{fu2024break} and online drafter adaptation to shifting input distributions~\citep{liu2023online}.
An orthogonal \textit{system optimization} amortizes verification cost by jointly checking multiple candidate continuations in a single target pass. SpecInfer~\citep{miao2024specinfer} introduces tree-structured speculation with shared-prefix tree attention; Sequoia~\citep{chen2024sequoia} casts tree topology as a constrained optimization that maximizes expected acceptance under a fixed verification budget; EAGLE-2~\citep{li2024eagle2} dynamically constructs context-dependent draft trees; and Medusa-2~\citep{cai2024medusa}
verifies candidates sampled from its multiple heads via tree attention.

\textbf{Diffusion Language Models.} Diffusion language models (dLLMs) cast text generation as iterative denoising over a masked sequence, allowing many positions to be decoded in parallel rather than strictly autoregressive. Recent efforts have substantially narrowed the quality gap with autoregressive models: LLaDA-1.5~\citep{zhu2025llada} introduces variance-reduced preference optimization tailored to discrete diffusion, and LLaDA-2.1~\citep{bie2026llada2} further accelerates sampling through token editing. Despite these advances, dLLMs still lag autoregressive counterparts in quality, limiting their adoption as standalone generators. This residual gap, however, is precisely what motivates their use as \emph{drafters} within speculative decoding: their parallel decoding can be paired with an autoregressive verifier that absorbs any residual quality loss while preserving quality.

\textbf{Speculative Decoding with Diffusion Draft Models.} A growing line of work replaces the sequential drafter with a diffusion-based alternative to overcome the linear drafting cost of autoregressive speculation. DiffuSpec~\citep{li2025diffuspec} adopts a large pre-trained dLLM as the drafter, achieving long acceptance lengths but incurring memory and latency overhead that partially offsets the speedup. PARD~\citep{an2025pard} mimics diffusion-style parallel generation with a small, low-cost autoregressive model, yet its limited capacity caps the attainable acceptance length. DFlash~\citep{chen2026dflash} introduces a lightweight block-diffusion drafter conditioned on target hidden features via KV injection and reports over $6\times$ lossless speedup; we adopt it as our first-stage drafter. Despite their parallelism, all of the above commit to a single draft sequence per verification step, so an early mismatch discards every downstream token, and the methods do not naturally interoperate with tree-based verification. A very recent concurrent work of~\cite{ringel2026accelerating} extends block-diffusion drafting to draft trees, but generates a large number of branches via uniform resampling, which dilutes the advantages of parallel verification on positions where the first draft was already correct.

\section{Conclusion}
We presented \name, a dual-diffusion-draft speculative decoding framework that converts a single block-diffusion draft into a structured set of shared-prefix candidates rather than committing to one sequence per verification. A first diffusion drafter emits a $\gamma$-token block together with per-position confidences; these confidences localize the most likely rejection boundary, from which the top-$K$ prefix lengths are selected; a variable-prefix diffusion drafter, trained with an anchor-weighted loss to support arbitrary prefix lengths, re-anchors at each selected prefix and proposes alternative continuations in one batched pass; the resulting $K + 1$ shared-prefix candidates are jointly verified by the target model. By construction, this targets the two failure modes that limit existing diffusion drafters---the \emph{scaling wall} of enlarging $\gamma$ and the \emph{error homogeneity} of resampling the same forward---while sidestepping the autoregressive drafting tax that bounds tree-based speculative decoders. The experiment results indicate that confidence-guided shared-prefix recovery---rather than block enlargement or uniform branch resampling---is the right primitive for combining parallel block drafting with the longest-correct-prefix verification rule.

\bibliographystyle{unsrt}
\bibliography{main}

\begin{thebibliography}{10}

\bibitem{liu2024deepseek}
Aixin Liu, Bei Feng, Bing Xue, Bingxuan Wang, Bochao Wu, Chengda Lu, Chenggang Zhao, Chengqi Deng, Chenyu Zhang, Chong Ruan, et~al.
\newblock Deepseek-v3 technical report.
\newblock {\em arXiv preprint arXiv:2412.19437}, 2024.

\bibitem{yang2025qwen3}
An~Yang, Anfeng Li, Baosong Yang, Beichen Zhang, Binyuan Hui, Bo~Zheng, Bowen Yu, Chang Gao, Chengen Huang, Chenxu Lv, et~al.
\newblock Qwen3 technical report.
\newblock {\em arXiv preprint arXiv:2505.09388}, 2025.

\bibitem{zeng2026glm}
Aohan Zeng, Xin Lv, Zhenyu Hou, Zhengxiao Du, Qinkai Zheng, Bin Chen, Da~Yin, Chendi Ge, Chenghua Huang, Chengxing Xie, et~al.
\newblock Glm-5: from vibe coding to agentic engineering.
\newblock {\em arXiv preprint arXiv:2602.15763}, 2026.

\bibitem{leviathan2023fast}
Yaniv Leviathan, Matan Kalman, and Yossi Matias.
\newblock Fast inference from transformers via speculative decoding.
\newblock In {\em International Conference on Machine Learning}, pages 19274--19286. PMLR, 2023.

\bibitem{chen2023accelerating}
Charlie Chen, Sebastian Borgeaud, Geoffrey Irving, Jean-Baptiste Lespiau, Laurent Sifre, and John Jumper.
\newblock Accelerating large language model decoding with speculative sampling.
\newblock {\em arXiv preprint arXiv:2302.01318}, 2023.

\bibitem{miao2024specinfer}
Xupeng Miao, Gabriele Oliaro, Zhihao Zhang, Xinhao Cheng, Zeyu Wang, Zhengxin Zhang, Rae Ying~Yee Wong, Alan Zhu, Lijie Yang, Xiaoxiang Shi, et~al.
\newblock Specinfer: Accelerating large language model serving with tree-based speculative inference and verification.
\newblock In {\em Proceedings of the 29th ACM International Conference on Architectural Support for Programming Languages and Operating Systems, Volume 3}, pages 932--949, 2024.

\bibitem{chen2024sequoia}
Zhuoming Chen, Avner May, Ruslan Svirschevski, Yuhsun Huang, Max Ryabinin, Zhihao Jia, and Beidi Chen.
\newblock Sequoia: Scalable, robust, and hardware-aware speculative decoding.
\newblock {\em arXiv preprint arXiv:2402.12374}, 2024.

\bibitem{li2025eagle3}
Yuhui Li, Fangyun Wei, Chao Zhang, and Hongyang Zhang.
\newblock Eagle-3: Scaling up inference acceleration of large language models via training-time test.
\newblock In {\em The Thirty-ninth Annual Conference on Neural Information Processing Systems}, 2025.

\bibitem{chen2026dflash}
Jian Chen, Yesheng Liang, and Zhijian Liu.
\newblock Dflash: Block diffusion for flash speculative decoding.
\newblock {\em arXiv preprint arXiv:2602.06036}, 2026.

\bibitem{pan2025fasttree}
Zaifeng Pan, Yitong Ding, Yue Guan, Zheng Wang, Zhongkai Yu, Xulong Tang, Yida Wang, and Yufei Ding.
\newblock Fasttree: Optimizing attention kernel and runtime for tree-structured llm inference.
\newblock {\em Proceedings of Machine Learning and Systems}, 7, 2025.

\bibitem{yaodeft}
Jinwei Yao, Kaiqi Chen, Kexun Zhang, Jiaxuan You, Binhang Yuan, Zeke Wang, and Tao Lin.
\newblock Deft: Decoding with flash tree-attention for efficient tree-structured llm inference.
\newblock In {\em 13th International Conference on Learning Representations, ICLR 2025}, pages 3587--3618. International Conference on Learning Representations, ICLR, 2025.

\bibitem{sadhukhan2024magicdec}
Ranajoy Sadhukhan, Jian Chen, Zhuoming Chen, Vashisth Tiwari, Ruihang Lai, Jinyuan Shi, Ian En-Hsu Yen, Avner May, Tianqi Chen, and Beidi Chen.
\newblock Magicdec: Breaking the latency-throughput tradeoff for long context generation with speculative decoding.
\newblock In {\em International Conference on Learning Representations (ICLR)}, 2025.

\bibitem{du2024glide}
Cunxiao Du, Jing Jiang, Xu~Yuanchen, Jiawei Wu, Sicheng Yu, Yongqi Li, Shenggui Li, Kai Xu, Liqiang Nie, Zhaopeng Tu, et~al.
\newblock Glide with a cape: A low-hassle method to accelerate speculative decoding.
\newblock In {\em International Conference on Machine Learning}, pages 11704--11720. PMLR, 2024.

\bibitem{li2024eagle2}
Yuhui Li, Fangyun Wei, Chao Zhang, and Hongyang Zhang.
\newblock Eagle-2: Faster inference of language models with dynamic draft trees.
\newblock In {\em Proceedings of the 2024 conference on empirical methods in natural language processing}, pages 7421--7432, 2024.

\bibitem{cai2024medusa}
Tianle Cai, Yuhong Li, Zhengyang Geng, Hongwu Peng, Jason~D Lee, Deming Chen, and Tri Dao.
\newblock Medusa: Simple llm inference acceleration framework with multiple decoding heads.
\newblock In {\em International Conference on Machine Learning}, pages 5209--5235. PMLR, 2024.

\bibitem{cobbe2021gsm8k}
Karl Cobbe, Vineet Kosaraju, Mohammad Bavarian, Mark Chen, Heewoo Jun, Lukasz Kaiser, Matthias Plappert, Jerry Tworek, Jacob Hilton, Reiichiro Nakano, et~al.
\newblock Training verifiers to solve math word problems.
\newblock {\em arXiv preprint arXiv:2110.14168}, 2021.

\bibitem{hendrycks2021math}
Dan Hendrycks, Collin Burns, Saurav Kadavath, Akul Arora, Steven Basart, Eric Tang, Dawn Song, and Jacob Steinhardt.
\newblock Measuring mathematical problem solving with the math dataset.
\newblock In J.~Vanschoren and S.~Yeung, editors, {\em Proceedings of the Neural Information Processing Systems Track on Datasets and Benchmarks}, volume~1, 2021.

\bibitem{maa2025aime}
{Mathematical Association of America}.
\newblock {American Invitational Mathematics Examination -- AIME 2025}.
\newblock \url{https://maa.org/maa-invitational-competitions}, February 2025.
\newblock Accessed: 2026-05-06.

\bibitem{chen2021humaneval}
Mark Chen, Jerry Tworek, Heewoo Jun, Qiming Yuan, Henrique Ponde De~Oliveira Pinto, Jared Kaplan, Harri Edwards, Yuri Burda, Nicholas Joseph, Greg Brockman, et~al.
\newblock Evaluating large language models trained on code.
\newblock {\em arXiv preprint arXiv:2107.03374}, 2021.

\bibitem{austin2021mbpp}
Jacob Austin, Augustus Odena, Maxwell Nye, Maarten Bosma, Henryk Michalewski, David Dohan, Ellen Jiang, Carrie Cai, Michael Terry, Quoc Le, et~al.
\newblock Program synthesis with large language models.
\newblock {\em arXiv preprint arXiv:2108.07732}, 2021.

\bibitem{jain2024livecodebench}
Naman Jain, Han, Alex Gu, Wen-Ding Li, Fanjia Yan, Tianjun Zhang, Sida Wang, Armando Solar-Lezama, Koushik Sen, and Ion Stoica.
\newblock Livecodebench: Holistic and contamination free evaluation of large language models for code.
\newblock 2025:58791--58831, 2025.

\bibitem{zheng2023mtbench}
Lianmin Zheng, Wei-Lin Chiang, Ying Sheng, Siyuan Zhuang, Zhanghao Wu, Yonghao Zhuang, Zi~Lin, Zhuohan Li, Dacheng Li, Eric Xing, et~al.
\newblock Judging llm-as-a-judge with mt-bench and chatbot arena.
\newblock {\em Advances in neural information processing systems}, 36:46595--46623, 2023.

\bibitem{taori2023alpaca}
Rohan Taori, Ishaan Gulrajani, Tianyi Zhang, Yann Dubois, Xuechen Li, Carlos Guestrin, Percy Liang, and Tatsunori~B. Hashimoto.
\newblock Stanford alpaca: An instruction-following llama model.
\newblock \url{https://github.com/tatsu-lab/stanford_alpaca}, 2023.

\bibitem{xu2024perfectblendredefiningrlhf}
Tengyu Xu, Eryk Helenowski, Karthik~Abinav Sankararaman, Di~Jin, Kaiyan Peng, Eric Han, Shaoliang Nie, Chen Zhu, Hejia Zhang, Wenxuan Zhou, et~al.
\newblock The perfect blend: Redefining rlhf with mixture of judges, 2024.

\bibitem{dao2024flashattention2}
Tri Dao.
\newblock Flashattention-2: Faster attention with better parallelism and work partitioning.
\newblock In {\em 12th International Conference on Learning Representations, ICLR 2024}, 2024.

\bibitem{ye2025flashinfer}
Zihao Ye, Lequn Chen, Ruihang Lai, Wuwei Lin, Yineng Zhang, Stephanie Wang, Tianqi Chen, Baris Kasikci, Vinod Grover, Arvind Krishnamurthy, et~al.
\newblock Flashinfer: Efficient and customizable attention engine for llm inference serving.
\newblock {\em Proceedings of Machine Learning and Systems}, 7, 2025.

\bibitem{li2024eagle}
Yuhui Li, Fangyun Wei, Chao Zhang, and Hongyang Zhang.
\newblock Eagle: speculative sampling requires rethinking feature uncertainty.
\newblock In {\em Proceedings of the 41st International Conference on Machine Learning}, pages 28935--28948, 2024.

\bibitem{fu2024break}
Yichao Fu, Peter Bailis, Ion Stoica, and Hao Zhang.
\newblock Break the sequential dependency of llm inference using lookahead decoding.
\newblock In {\em International Conference on Machine Learning}, pages 14060--14079. PMLR, 2024.

\bibitem{liu2023online}
Xiaoxuan Liu, Lanxiang Hu, Peter Bailis, Alvin Cheung, Zhijie Deng, Ion Stoica, and Hao Zhang.
\newblock Online speculative decoding.
\newblock {\em arXiv preprint arXiv:2310.07177}, 2023.

\bibitem{zhu2025llada}
Fengqi Zhu, Rongzhen Wang, Shen Nie, Xiaolu Zhang, Chunwei Wu, Jun Hu, Jun Zhou, Jianfei Chen, Yankai Lin, Ji-Rong Wen, et~al.
\newblock Llada 1.5: Variance-reduced preference optimization for large language diffusion models.
\newblock {\em arXiv preprint arXiv:2505.19223}, 2025.

\bibitem{bie2026llada2}
Tiwei Bie, Maosong Cao, Xiang Cao, Bingsen Chen, Fuyuan Chen, Kun Chen, Lun Du, Daozhuo Feng, Haibo Feng, Mingliang Gong, et~al.
\newblock Llada2. 1: Speeding up text diffusion via token editing.
\newblock {\em arXiv preprint arXiv:2602.08676}, 2026.

\bibitem{li2025diffuspec}
Guanghao Li, Zhihui Fu, Min Fang, Qibin Zhao, Ming Tang, Chun Yuan, and Jun Wang.
\newblock Diffuspec: Unlocking diffusion language models for speculative decoding.
\newblock {\em arXiv preprint arXiv:2510.02358}, 2025.

\bibitem{an2025pard}
Zihao An, Huajun Bai, Ziqiong Liu, Dong Li, and Emad Barsoum.
\newblock Pard: Accelerating llm inference with low-cost parallel draft model adaptation, 2025.

\bibitem{ringel2026accelerating}
Liran Ringel and Yaniv Romano.
\newblock Accelerating speculative decoding with block diffusion draft trees.
\newblock {\em arXiv preprint arXiv:2604.12989}, 2026.

\end{thebibliography}

\end{document}